\documentstyle[aps,preprint,pra,epsfig]{revtex}

\begin{document}

\draft

\title{Finite-Temperature Renormalization Group Analysis of 
Interaction Effects in $2D$ Lattices of Bose-Einstein Condensates}

\author{A. Smerzi$^{1,2}$, P. Sodano$^{3}$, and A. Trombettoni$^{4}$}
\address{$^1$ Istituto Nazionale per la Fisica della Materia BEC-CRS and
Dipartimento di Fisica, Universita' di Trento, I-38050 Povo, Italy\\
$^2$ Theoretical Division, Los Alamos
National Laboratory, Los Alamos, NM 87545, USA\\
$^3$ Dipartimento di Fisica and Sezione I.N.F.N., Universit\`a di 
Perugia, Via A. Pascoli, I-06123 Perugia, Italy\\
$^4$ Istituto Nazionale per la Fisica della Materia and
Dipartimento di Fisica, Universita' di Parma,
parco Area delle Scienze 7A, I-43100 Parma, Italy\\
}

\date{\today}
\maketitle

\begin{abstract}
By using a renormalization group analysis, 
we study the effect of interparticle interactions on the 
critical temperature $T_{BKT}$ at which the Berezinskii-Kosterlitz-Thouless 
(BKT) transition occurs for Bose-Einstein condensates loaded at 
finite temperature in a $2D$ optical lattice. 
We find that $T_{BKT}$ decreases as the 
interaction energy decreases; when $U/J=36/\pi$ one has $T_{BKT}=0$, 
signaling the possibility of a quantum phase transition of BKT type.
\end{abstract}


\section{Introduction}

It has been recently suggested \cite{trombettoni03} 
that a $2D$ optical lattice of Bose-Einstein 
condensates \cite{greiner01} may allow for the observation 
of a finite-temperature phase transition to a superfluid regime 
where the phases of the single-well condensates are coherently aligned. 
In fact, in an appropriate range of parameters, 
the thermodynamical properties of the bosonic lattice at finite temperature 
may be well described by the Hamiltonian of the $XY$ model 
\cite{trombettoni03}, and, as it is well known, the $2D$ $XY$ model 
exhibits the Berezinskii-Kosterlitz-Thouless (BKT) 
transition \cite{BKT,minnaghen87,kadanoff00}. 
The BKT phase transition occurs via an 
unbinding of vortex defects: the low-temperature phase, 
for $T$ below the BKT temperature $T_{BKT}$,  
is characterized by the presence of bound vortex-antivortex pairs 
and the spatial correlations exhibit a power-law 
decay. For $T \sim T_{BKT}$ the pairs starts to unbind 
(see e.g. \cite{tobochnik79,leoncini98}): in the high-temperature phase, 
only free vortices are present, leading to an
effective randomization of the phases and to an exponential decay 
of the correlation functions.

The Hamiltonian describing the properties of bosons 
in deep optical lattices 
is the so called Bose-Hubbard Hamiltonian \cite{jaksch98}, 
in which two terms are present: a kinetic term 
describing the hopping of the bosons with 
tunneling rate $t$, and a potential term describing 
the interaction between bosons in the wells of the array 
with interaction energy $U$ (which is proportional to the 
$s$-wave scattering length). The description of bosons in 
deep lattices by means of the Bose-Hubbard Hamiltonian holds 
also at finite temperature, provided that the temperature effects 
do not induce the occupation of higher bands: for a $2D$ optical 
lattice confined to a plane by a magnetic potential having frequency 
$\omega_z$, this implies that the Bose-Hubbard Hamiltonian 
description is 
valid up temperatures $T$ such that $\hbar \omega_z \gtrsim k_B T$.   
When the average number of bosons $N_0$ per site is high enough 
and the interaction energy is larger than $t/N_0$, one may map - 
{\em also at finite temperature} - the Bose-Hubbard Hamiltonian 
in the quantum phase Hamiltonian \cite{trombettoni03}: 
the conjugate variables are the phases and the particle numbers 
of the condensates in the different sites of the lattice. 
Thus, also at finite temperature, the phase diagram is 
determined by the competition of two energies: 
the Josephson energy $J \approx 2 t N_0$, 
proportional to the tunneling rate between neighbouring sites of the $2D$ 
square optical lattice, and the interaction energy $U$. For $U=0$, the 
quantum phase Hamiltonian reduces to the $XY$ Hamiltonian, exhibiting then 
a BKT transition at $T_{BKT} \sim J/k_B$: for $T < T_{BKT}$ the system 
as a whole behaves as a superfluid with phase coherence across the array.  
Very accurate Monte Carlo simulations yield, in the thermodynamic limit, 
$T_{BKT} = 0.898 J/k_B$ \cite{gupta88}. 

In this paper we determine the effect of the 
interaction energy $U$ on the BKT transition. 
When $U \ll J$, a BKT transition still occurs at a critical 
temperature $T_{BKT}(U)$: we find that  
$T_{BKT}(U)$ is smaller than $T_{BKT}$ and decreases with $U$ 
(see Fig. 1). Intuitively speaking, the quantum phase Hamiltonian, 
as well as the Bose-Hubbard Hamiltonian, in the limit $U \gg J$ 
describes a Mott insulator, while in 
the opposite limit, $J \gg U$, the array behaves as a 
phase-coherent superfluid 
(e.g., see \cite{sondhi97,fazio01}): thus, when $U$ increases, 
the superfluid region in the phase diagram decreases. 
Although in two dimensions there is not long-range order at finite temperature 
due to the Mermin-Wagner theorem, 
still the system exhibits superfluid behavior 
for $T < T_{BKT}$ \cite{fazio01}: thus one expects  
that $T_{BKT}(U)$ should decrease when $U$ increases. 
We also find (see Section III) that, when $U/J$ is equal to the 
critical value $(U/J)_{cr}=36/\pi$, 
one has that $T_{BKT}(U) = 0$, signaling the possibility of a ($T=0$) 
quantum phase transition. 

The quantum phase Hamiltonian describes also the behavior 
of superconducting Josephson networks below the temperature 
$T_{BCS}$ at which each junction becomes 
superconducting \cite{fazio01,simanek94,martinoli00}. 
The momenta conjugate to the phases of the macroscopic 
wavefunctions of the superconducting grains are 
the number of Cooper pairs, $J$ is the Josephson energy of the 
superconducting Josephson junctions and $U$ is the charging energy 
due to the Coulomb repulsion between Cooper pairs. 
We observe that, in superconducting 
Josephson arrays, the interaction term is written 
in general as $\sum_{ij} U_{ij} N_i N_j$ where $N_i$ is the Cooper pairs 
number at the site $i$: $U_{ij}$ is proportional to the inverse 
of the capacitance matrix and may be also non-diagonal (i.e., $U_{ij} 
\neq 0$ with $i \neq j$). At variance, for bosons in deep optical lattices, 
the interaction term is $\sum_{i} U_{ii} N_i^2$, which corresponds, 
in a suitable range of parameters, to a $diagonal$ quantum phase model. 

The analogy between superconducting Josephson networks and 
atomic gas in deep optical lattices is then clear: 
in each well of the periodic potential there is a condensate grain, 
appearing at the Bose-Einstein condensation temperature 
$T_{BEC}$. When $T_{BEC}$ is larger than all other energy scales, 
the atoms in the well $i$ of the $2D$ optical lattice may be 
described by a macroscopic wavefunction. 
It becomes apparent, then, that an optical network 
can be regarded as a network of {\em bosonic} Josephson junctions. 
Furthermore, in both systems, when the interaction term is neglected 
respect to the energy associated to the particle hopping, one has 
that the whole $2D$ array becomes superfluid at the temperature 
$T_{BKT}$ at which the BKT transition occurs, with $T_{BKT}$ smaller 
than $T_{BCS}$ for superconducting Josephson 
networks and smaller than $T_{BEC}$ for bosonic Josephson networks.

The plan of the paper is the following: in Section II we introduce the 
effective Hamiltonian describing bosons hopping on a deep optical lattice 
and, by using a semiclassical approximation 
\cite{jose84}, we compute the effective Josephson 
energy in presence of the interaction energy $U$.
In Section III we evaluate the effect of the quantum fluctuations 
on $T_{BKT}$ by putting in the renormalization group equations 
the effective Josephson energy obtained in Section II; 
a comparison with previous results is then carried out. 
Details of the computations skipped in the text are in the 
Appendix A. Section IV is devoted to some concluding remarks. 

\section{The renormalized Josephson energy}

$2D$ optical lattices are created using two standing waves \cite{greiner01}: 
when the polarization vectors of the two laser fields are orthogonal, 
the periodic potential is 
\begin{equation}
V_{opt}=V_0 [ \sin^2{(kx)} + \sin^2{(ky)} ],
\label{V_opt}
\end{equation}
where $k=2 \pi/\lambda$ is the wavevector of the lattice beams. 
$V_0$ is usually expressed in units of $E_R=\hbar^2 k^2/2m$ 
(where $m$ is the atomic mass): 
in \cite{greiner01} it is $\lambda=852 \, nm$ and 
$E_R=h \cdot 3.14 kHz$. 
Around the minima of the potential (\ref{V_opt}) one has 
$V_{opt} \approx m\tilde{\omega}_r^2 (x^2 + y^2)/2$ 
with 
\begin{equation}
\tilde{\omega}_r=\sqrt{\frac{2 V_0 k^2} {m}}.
\label{omega_tilde}
\end{equation} 
When $\tilde{\omega}_r \gg \omega_z$ (with $\omega_z$ 
the frequency of the confining magnetic potential superimposed 
to the optical potential and acting along $z$), 
the system realizes a square array of tubes, 
i.e. an array of harmonic traps elongated along the $z$-axis 
\cite{greiner01}. 

When all the relevant energy scales are small compared to the excitation 
energies, one can expand the field operator \cite{jaksch98} as 
\begin{equation}
\hat{\Psi}(\vec{r},t)= \sum_i \hat{\psi}_i(t) 
\Phi_i(\vec{r})
\label{expansion} 
\end{equation}
with $\Phi_i(\vec{r})$ the Wannier wavefunction localized 
in the $i$-th well 
(normalized to $1$) and $\hat{N}_i=\hat{\psi}^{\dag}_i 
\hat{\psi}_i$ the bosonic number operator. 
Substituting the expansion of $\hat{\Psi}(\vec{r},t)$ 
in the full quantum Hamiltonian, one gets  
the effective Hamiltonian describing the bosons hopping on 
the deep optical lattice \cite{jaksch98,fisher89} 
\begin{equation}
H=-t \sum_{<i, j>} 
(\hat{\psi}^{\dag}_i \hat{\psi}_j+ h.c.)
+ \frac{U}{2} \sum_i \hat{N}_i (\hat{N}_i-1).
\label{B-H}
\end{equation}
In Eq.(\ref{B-H}) $\sum_{<i, j>}$ denotes a sum over all the 
distinct pairs of nearest neighbours; 
$t$ and $U$ are respectively the tunneling rate and 
the interaction energy, and are given by 
\begin{equation}
t \simeq - \int d\vec{r} \, \Biggl[ \frac{\hbar^2}{2m} 
\vec{\nabla} \Phi_i \cdot \vec{\nabla} \Phi_j + 
\Phi_i  V_{ext} \Phi_j \Biggr] 
\label{t}
\end{equation}  
and 
\begin{equation}
U = \frac{4 \pi \hbar^2 a}{m} \int d\vec{r} \, 
\Phi_i^4
\label{U} 
\end{equation}
($a$ is the $s$-wave scattering length).

As discussed in the Introduction, one can show that, 
when $V_0$ and $\omega_z$ are large enough, 
the system is described by the Hamiltonian (\ref{B-H}) up to temperatures 
$T \sim \hbar \omega_z / k_B$ \cite{trombettoni03}. 
Upon defining $J \approx 2 t N_0$, the Hamiltonian (\ref{B-H}),  
for $N_0 \gg 1 $ and $J/N_0^2 \ll U$ \cite{anglin01}, reduces to 
\begin{equation}
\hat{H}=- J \sum_{<i,j>} \cos{(\theta_i-\theta_j)}
- {U \over 2} \sum_i \frac{\partial^2}{\partial \theta_i^2}.  
\label{Q-P-M}
\end{equation}
For a $2D$ lattice with $V_0$ between $20E_R$ and $25E_R$ and 
$N_0 \approx 170$ as in \cite{greiner01}, one sees 
that the condition $\gg J/N_0^2$ is rather well 
satisfied and that $J/k_B$ is of order of $20nK$. 
For $U=0$, (\ref{Q-P-M}) is the $XY$ Hamiltonian:
\begin{equation}
\hat{H}=- J \sum_{<i,j>} 
\cos{(\theta_i-\theta_j)}.
\label{X-Y}
\end{equation}

The Hamiltonian (\ref{Q-P-M}) describes the so-called 
quantum phase model \cite{fazio01,simanek94,martinoli00}. 
There is an huge amount of literature on the properties of $2D$ 
superconducting Josephson arrays studied by means of the quantum phase model 
\cite{fazio01,simanek94,martinoli00}. A (not exhaustive) list of 
relevant papers includes mean-field and coarse-graining 
approaches \cite{vanotterlo93,kopec00,grignani00}, 
Monte Carlo results 
\cite{jacobs84,rojas96,capriotti03,alet03,alsaidi04}, 
renormalization group calculations \cite{jose84,rojas96} 
and self-consistent harmonic approximations \cite{rojas96,kim97,cuccoli00} 
(more references are in \cite{fazio01}). In the following we shall study 
the renormalization-group equations in which it is used 
an effective value of the Josephson energy computed within the 
harmonic approximation \cite{jose84}. 

The starting point is the partition function $Z$ of the quantum phase model: 
using the path integral formalism, 
from (\ref{Q-P-M}), one has 
\begin{equation}
Z=\int {\cal D} \theta e^{-\frac{1}{\hbar} S[\theta]},
\label{part_funct}
\end{equation}  
where the action $S$ is given by
\begin{equation}
S[\theta]=
\int\limits_{0}^{\hbar \beta} {d\tau \Biggl[\frac{\hbar^2}{2U} 
\sum\limits_j { \Biggl(\frac{\partial \theta_j}{\partial
\tau}\Biggr)^2}+ J \sum\limits_{<i, j>} {(1-
\cos\theta_{ij})} \Biggr]}
\label{azione}
\end{equation}
with $\beta=1/k_B T$ and 
$\theta_{ij} \equiv \theta_{i}-\theta_{j}$.
Separating the phases as $\theta_i(\tau)=\varphi_i+
\delta_i(\tau)$, where $\varphi_i$ is a
static vortex configuration and $\delta_i(\tau)$ is a fluctuation
about $\varphi_i$, the path-integral partition function 
(\ref{part_funct}) can be written as 
\begin{equation}
Z=\int {\cal D} \varphi {\cal D} \delta e^{-\frac{1}{\hbar} 
S[\varphi,\delta]}
\label{ZETA} 
\end{equation}
where 
\begin{equation}
S[\varphi,\delta]=\int\limits_{0}^{\hbar \beta} 
{d\tau \Biggl[ \frac{\hbar^2}{2U} \sum\limits_{i} 
{\dot{\delta}_i^2 (\tau)}+J \sum\limits_{<i,j>} 
{(1- \cos\varphi_{ij} \, \cos\delta_{ij})} \Biggr]},
\label{azione_phi_delta}
\end{equation}
with $\varphi_{ij} \equiv \varphi_{i}-\varphi_{j}$ e 
$\delta_{ij} \equiv \delta_{i}-\delta_{j}$. 
Assuming that $\varphi_i$ and $\delta_i$ are slowly varying over the 
size of the array \cite{jose84,simanek94}, i.e. 
\begin{equation}
\cos\varphi_{ij} \approx 1-\varphi_{ij}^2/2
\label{APPR1}
\end{equation}
and 
\begin{equation}
\cos\delta_{ij} \approx 1-\delta_{ij}^2/2,
\label{APPR2}
\end{equation}
one gets 
\begin{equation}
Z=Z_0 \int{{\cal D}\varphi \exp\Bigl\{-{1 \over 2} 
\beta \bar{J} \sum\limits_{<i,j>} {\varphi_{ij}^2}\Bigr\}},
\label{part_funct_phi_delta}
\end{equation} 
where $Z_0 \equiv \int{{\cal D}\delta \, e^{-{1 \over \hbar} S_0[\delta]}}$ 
and 
\begin{equation}
S_0[\delta]=\int\limits_{0}^{\hbar \beta} 
{d\tau \Biggl[\frac{\hbar^2}{2U} \sum\limits_{i} {\dot{\delta}_i^2}+
{J \over 2} \sum\limits_{<i,j>} {\delta_{ij}^2}\Biggr]}.
\end{equation} 
In Eq.(\ref{part_funct_phi_delta}) 
$\bar{J}$ is the renormalized Josephson energy, which is given by
\begin{equation}
\bar{J} \simeq J \Bigl(1-{1 \over 2} <\delta_{ij}^2>_0\Bigr)
\label{J_rin}
\end{equation} 
with 
\begin{equation}
<\delta_{ij}^2>_0 \equiv \frac{1}{Z_0} 
\int{{\cal D}\delta \, e^{-{1 \over \hbar} S_0[\delta]}\delta_{ij}^2}.
\label{delta_ij}
\end{equation} 
The evaluation of $<\delta_{ij}^2>_0$ can be carried out in a standard way  
\cite{simanek94} and one has
\begin{equation}
<\delta_{ij}^2>_0=\sqrt{\frac{\pi U}{J}} 
\frac{1}{\eta^3} \int\limits_{0}^{\eta} x^2 \coth{x} \, dx
\label{delta}
\end{equation}
where 
\begin{equation}
\eta=\beta \sqrt{\pi U J}.
\label{eta}
\end{equation} 

\section{Renormalization group equations}

In the renormalization group equations for the $2D$ $XY$ model 
\cite{BKT,minnaghen87,kadanoff00} the scale-dependent screened charge 
${\cal K}$ depends on the dimensionless scaling parameter 
$l=\log{(r/a)}$ (where $r$ is the vortex
distance and $a$ is the lattice spacing) and it is given by 
${\cal K}(l)=\beta J/\epsilon(l)$, where the 
dielectric constant $\epsilon(l)$ expresses the screening of the 
vortex-antivortex interaction due to the presence of other vortices 
\cite{kadanoff00}. The renormalization group recursion equations 
read \cite{kadanoff00}
\begin{equation}
\frac{d {\cal K}^{-1}(l)}{dl}=4 \pi^3 y^2(l)
\label{prima_rec}
\end{equation}
and 
\begin{equation}
\frac{d y(l)}{dl}=[2- \pi {\cal K}(l)] y(l)
\label{seconda_rec}
\end{equation}
where $y \propto r^2 e^{-\beta V(r)/2} e^{-\beta \mu}$, 
with $e^{-\beta \mu}$ is the fugacity for creating a vortex pair 
and $V(r)$ correspond to the screened vortex-antivortex potential. 
The study of the scaling equations (\ref{prima_rec})-(\ref{seconda_rec}) 
about the fixed point $y_f=0$, ${\cal K}_f=2/\pi$ shows that 
the BKT transition occurs at $2-\pi {\cal K}(l=0) \approx 0$, where 
${\cal K}(l=0)=\beta J/\epsilon(l=0)=\beta J$ \cite{kadanoff00}.

In presence of the quantum fluctuations, one has to replace the 
scale-dependent charge ${\cal K}(l)$ by 
$\bar{\cal K}(l)$ with $\bar{\cal K}(l=0)= \beta \bar{J}$ and  
$\bar{J}$ given by Eqs.(\ref{J_rin}) and (\ref{delta}). 
Denoting with $T_{BKT} (U)$ 
the BKT transition temperature for a given $U$, one finds  
the following equation for $K \equiv J/ k_B T_{BKT}(U)$:
\begin{equation}
F(K)=2-\pi K  \Biggl(1-\frac{1}{2 \pi^2 X_u K^3} \int_{0}^{\pi \sqrt{X_u} K}
x^2 \coth{x} \, dx \Biggr)=0
\label{radice}
\end{equation}
where $X_u=U/\pi J$.

The root of Eq.(\ref{radice}) indicates the critical point 
at which a BKT transition occurs. The eventual occurrence 
of a double root for Eq.(\ref{radice}) might correspond to what is called 
in literature a {\em reentrant} behavior. Indeed it has been 
often argued (see Refs. in the review 
\cite{fazio01}) that the quantum phase model 
may undergo at low temperatures a reentrant transition 
induced by the quantum fluctuations: namely, fixing $U/J$ and 
lowering the temperature, one could switch from
an insulating phase to a superconducting one at $T_{BKT}(U)$ and then, 
lowering further the temperature, one finds another critical temperature 
$T^{(1)}(U)$ at which the system comes back ({\em reenters}) in the 
insulating phase. Consistent with the reentrant scenario is the dramatic 
decrease of the specific heat at very low temperatures  
\cite{jacobs84} and, as discussed, 
the presence of double roots for $T_{BKT}$ in the 
renormalization group equations. 

The phenomenon of the reentrance - although not universal - 
is a non-perturbative effect: in fact, opposite 
results may be obtained by means of 
different truncations for the series expansion of the function $F$. 
For instance, if in Eq.(\ref{radice}) one expands the hyperbolic cotangent 
as $\coth x \approx 1/x+x/3$, one gets an expression for $F(K)$ 
which gives two roots for $T_{BKT}$. If instead, for $\eta<\pi$, 
one uses the expansion $\coth x=\frac{1}{x}+\sum\limits_{n=1}^{\infty} 
\frac{2^{2n} B_{2n}}{(2n)!} x^{2n-1}$ where the $B_n$'s are the 
Bernoulli numbers \cite{abramowitz72}, one obtains the equation: 
\begin{equation}
F(K)=2-\pi K \Biggl(1-\frac{1}{4 K}-\sum\limits_{n=1}^{\infty} 
\frac{6^n \,  2^{4n} \, B_{2n} \, x_u^n \, K^{2n-1}}
{2 \, (2n+2) \, (2n)!} \Biggr)=0
\label{bernoulli}
\end{equation}  
with $x_u=\pi U/24 J$. Since $B_{4n}<0$ for $n=1,2, \ldots$, and 
$B_{4n+2}>0$ for $n=0,1,2, \ldots$, if one truncates the sum 
in Eq.(\ref{bernoulli}) to the $(2n+1)$-th order with 
$n=0,1,2,\ldots$, one finds 
$F(K) \to \infty$ for $K \to \infty$ and 
then Eq.(\ref{bernoulli}) has two roots; 
at variance, if one truncates the sum 
to the $(2n)$-th order with 
$n=1,2,\ldots$, one finds 
$F(K) \to -\infty$ for $K \to \infty$ and 
Eq.(\ref{bernoulli}) admits only one root. Thus one cannot truncate 
the sum in Eq.(\ref{bernoulli}) to any finite order, even for $U/J \ll 1$; 
the origin of this problem is that the expansion used 
to get (\ref{bernoulli}) is a series with terms having alternating 
signs. In addition, to find the roots 
of the equation $F(K)=0$ one has to evaluate the integral 
in (\ref{delta}) up to $\sqrt{UJ}/k_B T_{BKT}$, and, also 
with $U$ small, $\eta$ may become large requiring to use all the orders 
in Eq.(\ref{bernoulli}). 

A way to overcome the above mentioned 
difficulties is to use in Eq.(\ref{radice}) 
the Mittag-Leffler expansion of 
the hyperbolic cotangent, i.e. 
$\pi \coth \pi x=1/x+2x\sum\limits_{n=1}^{\infty} (x^2+n^2)^{-1}$: 
in this way Eq.(\ref{radice}) can be written as 
\begin{equation}
F(K)=2-\pi K  \Biggl[1-\frac{1}{4 K}-\frac{1}{X_u 
K^3} \, g(K,X_u) \Biggr]=0
\label{radice_impr}
\end{equation}
where
\begin{equation}
g(K,X_u) \equiv \sum\limits_{n=1}^{\infty} 
\Bigl[\frac{X_u K^2}{2}-\frac{n^2}{2} 
\log \Bigl(1+ \frac{X_u K^2}{n^2}\Bigr)\Bigr].
\label{g}
\end{equation}
An analytic expression for $F(K)$ valid for $U/J \ll 1$ is given 
by Eq.(\ref{approxx}) in the Appendix A.

A detailed study of the function $F(K)$ is 
in the Appendix A. Here we observe that, in the limit $U/J \to 0$, 
one finds $2+\frac{\pi}{4}-\pi K=0$, from which
$K_0 \equiv K(U/J \to 0)=\frac{8+\pi}{4 \pi}$ and 
$k_B T_{BKT} (0) \simeq 1.128 J$. Furthermore, 
for $U/J<36/\pi$, one can show (see the Appendix A) that 
$F(K) \to - \infty$ for $K \to \infty$ and that $F^\prime (K)<0$: 
since $F$ tends to the positive value $2+\pi/4$ for $K \to 0$, 
one can safely conclude 
that, for $U/J<36/\pi$, Eq.(\ref{radice_impr}) has an {\it unique} solution. 
At variance, one can show that, for $U/J>36/\pi$, Eq.(\ref{radice_impr}) 
does not admit any solution, while for 
$U/J=36/\pi$ one has $K \to \infty$, i.e. $T_{BKT} (U) \to 0$. 
A plot of $F(K)$ and $F^\prime(K)$ for three values of $U/J$ 
respectively smaller than, equal to and larger than $36/\pi$ is 
given in Figs. 2 and 3.
One may infer that, at $T=0$, a phase transition is expected to occur 
at the critical value $(U/J)_{cr}=36/\pi$: this 
value is in reasonable agreement with the 
mean-field estimates for the $T=0$ transition. In fact, 
for the diagonal quantum phase model, one has that 
the mean-field prediction 
\cite{vanotterlo93,grignani00} 
is $(U/J)_{cr} \approx 2z =8$ where $z=4$ is the number 
of nearest neighbours. 

In Fig. 1 we plot as empty circles the values of $T_{BKT}(U)/T_{BKT}(0)$ 
as a function of $U/J$ from the numerical solution 
of Eq.(\ref{radice_impr}). 
If one uses the analytic expression (\ref{approxx}) 
for the function $F(K)$ one gets, for $U/J \lesssim 1/2$, 
an error lesser than $1 \%$. A much better estimate may be obtained 
by expanding the function 
$F(K)$ near $K_0$: in this way Eq.(\ref{radice_impr}) reads 
\begin{displaymath}
F(K) \approx F(K_0)+ (K-K_0) \cdot 
F^{\prime}(K_0)=0,
\end{displaymath}
and then
\begin{equation}
K \approx K_0 - \frac{F(K_0)}
{F^{\prime}(K_0)}.
\label{sviluppo}
\end{equation}
In Fig. 1 Eq.(\ref{sviluppo}) is plotted as a solid line.

We may conclude that Eq.(\ref{radice_impr}) admits, for small $U/J$, 
a unique solution and this excludes the possibility of reentrant 
behavior. Of course, our conclusion 
relies on the approximations (\ref{APPR1})-(\ref{APPR2}) 
made in order to estimate $\bar{J}$ and $<\delta_{ij}^2>_0$. 
A more careful treatment accounting for the 
periodicity of the phases in Eq.(\ref{azione_phi_delta}) 
seems to be highly desirable to put on a more solid base 
all the contrasting issues related to the reentrant 
behavior of the systems described by the Hamiltonians (\ref{B-H}) and 
(\ref{Q-P-M}). It is comforting to see that the experimental study  
of weakly interacting bosons on $2D$ optical networks and 
the investigations of their superfluidity properties at very low temperatures 
could provide new insights also on this 
intriguing and yet poorly understood problem 
(see also the recent papers \cite{kleinert03}).

\section{Concluding remarks}
  
In this paper we studied, by means of a renormalization group analysis, 
the effect of interparticle interactions on the 
critical temperature at which the Berezinskii-Kosterlitz-Thouless 
transition occurs for Bose-Einstein condensates loaded in a $2D$ 
optical lattice at finite temperature. We determined the shift 
of the Berezinskii-Kosterlitz-Thouless transition temperature 
induced by the interaction term and we compared our findings with 
previously known results.  

\appendix 

\section{Properties of the renormalization group equation}

In this Appendix we study the properties of the function 
$F(K)$ defined in Eq.(\ref{radice_impr}). 
We show in (1) that $F(0)>0$, in (2) that 
$F(K) \to -\infty$ for $K \to \infty$ when $U/J<36/\pi$ and 
in (3) that for $U/J<36/\pi$ one has $F^\prime(K)<0$: 
one may then conclude that the equation $F(K)=0$ has only 
one root for $U/J<36/\pi$. Finally, we derive an analytic expression 
for $F(K)$ holding for small $U/J$. 

From Eq.(\ref{radice_impr}) one can see that:\\
(1) for $U/J \to 0$ one has $2+\frac{\pi}{4}-\pi K=0$, 
from which $K=\frac{8+\pi}{4 \pi}$ and $K_B T_{BKT} \simeq 1.128 J$: 
indeed, for small $U/J$, one has that
\begin{displaymath}
\frac{X_u K^2}{2}- \frac{n^2}{2} \log 
\Bigl(1+\frac{X_u K^2}{n^2}\Bigr)=\frac{X_u K^2}{2}- 
\frac{n^2}{2} \Bigl(\frac{X_u K^2}{n^2}-\frac{X_u^2 K^4}{2 n^4}+
 \ldots\Bigr)=\frac{X_u^2 K^4}{4 n^2}-\ldots
\end{displaymath}
and thus 
\begin{displaymath}
\frac{1}{X_u K^2} g(K,X_u) \approx \frac{1}{X_u K^2} 
\sum\limits_{n=1}^{\infty} \frac{X_u^2 K^4}{4 n^2} \to 0.
\end{displaymath}
In the same way, $F(K) \to 2+\frac{\pi}{4}$ for $K \to 0$.\\

(2) $F(K) \to -\infty$ per $K \to \infty$ for $x_u<3/2$ 
(where $x_u=\pi^2 X_u/24=\pi U/24J$): indeed for large $K$ 
\begin{displaymath}
F(K) \approx 2-\pi K \Bigl(1-\frac{1}{48 x_u K^3} \int\limits_{0}^{2 \sqrt{6} 
\sqrt{x_u} K} dx x^2 \Bigr)=2-\pi K \Bigl(1-\frac{\sqrt{6}}{3} 
\sqrt{x_u}\Bigr):
\end{displaymath}
then, for $K \to \infty$, $F(K) \to -\infty$ with $x_u<3/2$ 
and $F(K) \to \infty$ with $x_u>3/2$.\\ 

(3) For $x_u<3/2$, it is $F^\prime (K)<0$ for $K>0$: 
indeed 
\begin{equation}
F^\prime(K)=-\pi-\frac{2 \pi}{X_u K^3} \, g(K,X_u)+\frac{\pi^2 X_u^{1/2}}
{2} \Bigl[\coth(\pi X_u^{1/2} K)- \frac{1}{\pi X_u^{1/2} K}\Bigr].
\label{21}
\end{equation}
Since $\lim_{x \to 0} \Bigl(\coth x-\frac{1}{x}\Bigr)=0$, 
one gets
\begin{displaymath}
\lim_{K \to 0} F^\prime(K)=-\pi.
\end{displaymath}
For large $K$ one finds
\begin{displaymath}
F^\prime(K) \approx -\pi+\frac{\pi^2}{6} X_u^{1/2}.
\end{displaymath}
Therefore one has that $\lim_{K \to \infty} F^\prime(K)<0$ for 
$X_u<36/\pi^2$, and $\lim_{K \to \infty} F^\prime(K)>0$ for 
$X_u>36/\pi^2$. 
Using similar arguments, one can show that, for $K>0$, it is 
\begin{equation}
F^\prime(K)<-\pi+\frac{\pi^2}{6} X_u^{1/2}.
\label{21d}
\end{equation}
which implies that, for $X_u<36/\pi^2$, one has $F^\prime(K)<0$. 
The behavior of $F(K)$ and $F^\prime(K)$ for three values of $U/J$ 
respectively smaller than, equal to and larger than $36/\pi$ is 
plotted in Figs. 2 and 3.

One may also obtain an analytic approximation for $F(K)$ holding 
for $U/J \ll 1$ by putting 
$z=\sqrt{X_u} K$ and using $\log (1+z)=\sum\limits_{j=1}^{\infty} 
\frac{(-1)^{j+1} z^j}{j}$ for $\mid z \mid<1$: one has 
\begin{equation}
g(K,X_u)=\sum\limits_{j=2}^{\infty} \frac{(-1)^j z^{2j}}{2j} \zeta(2j-2)
\label{appross}
\end{equation}
where $\zeta (j)=\sum\limits_{n=1}^{\infty} \frac{1}{n^j}$ is the
Riemann zeta-function. One finds
\begin{equation}
\sum\limits_{j=2}^{\infty} \frac{(-1)^j z^{2j}}{2j} \zeta(2j)=
\frac{1}{2} \Bigl\{\frac{\pi^2 z^2}{6}+\log \frac{\pi z}{\sinh \pi z}\Bigr\}.
\label{appross1}
\end{equation} 
For large $n$ one has the recurrence relation \cite{abramowitz72}
\begin{displaymath}
\zeta(n+1) \simeq \frac{1}{2} [1+ \zeta(n)]. 
\end{displaymath}
Substituting back in Eq.(\ref{appross}) and using (\ref{appross1}), 
one has
\begin{displaymath}
g(K,X_u) \approx X_u K^2 \Bigl(\frac{\pi^2}{3}-{3 \over 2}\Bigr)+
2 \log \frac{\pi X_u^{1/2} K}{\sinh \pi X_u^{1/2} K}+
{3 \over 2} \log (1+X_u K^2)
\end{displaymath}
and finally 
\begin{equation}
F \approx 2-\pi K 
\Bigg[1-\frac{1}{4 K}-\frac{1}{X_u 
K^3} \Bigg( X_u K^2 \bigg(\frac{\pi^2}{3}-{3 \over 2}\bigg)+
2 \log \frac{\pi X_u^{1/2} K}{\sinh \pi X_u^{1/2} K}+
{3 \over 2} \log (1+X_u K^2) \Bigg) \Bigg].
\label{approxx}
\end{equation}


\begin{figure}
\centerline{\psfig{figure=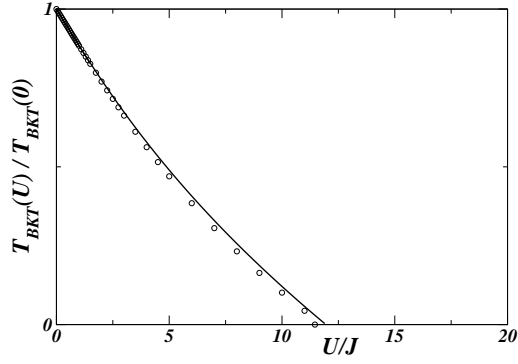,width=60mm,angle=270}}
\caption{The BKT critical temperature $T_{BKT}(U)$ 
(in units of $T_{BKT}(0)$) as a function  
of $U/J$. Empty circles: numerical solution of Eq.(\ref{radice_impr}); 
solid line: Eq.(\ref{sviluppo}).}
\label{fig1}
\end{figure}

\begin{figure}
\centerline{\psfig{figure=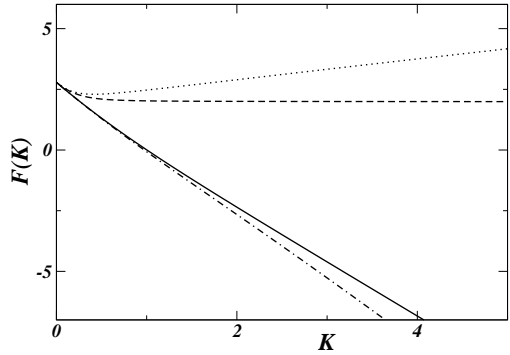,width=60mm,angle=270}}
\caption{$F(K)$ for $U/J=1$ (solid line), $36/\pi$ (dashed line) 
and $15$ (dotted line). The dot-dashed line is the analytic approximation 
(\ref{approxx}) for $F$ holding for $U/J \ll 1$. For 
$U/J=36/\pi$ $F$ asymptotically tends to $0$.}
\label{fig2}
\end{figure}

\begin{figure}
\centerline{\psfig{figure=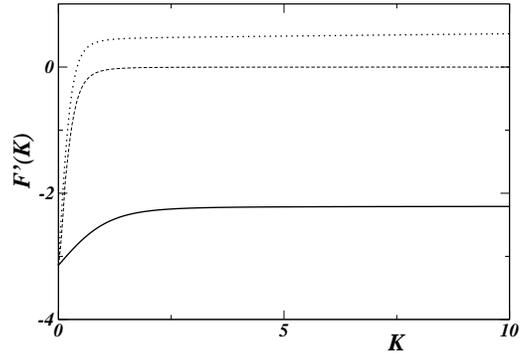,width=60mm,angle=270}}
\caption{$F^\prime(K)$ 
for $U/J=1$ (solid line), $36/\pi$ (dashed line) 
and $15$ (dotted line).}
\label{fig3}
\end{figure}

\end{document}